\documentclass[aps,prx,twocolumn,amsmath,amssymb,superscriptaddress,letterpaper]{revtex4}
\usepackage{times}
\usepackage{amsfonts}
\usepackage{mathrsfs}
\usepackage[version=3]{mhchem}
\usepackage{graphicx}% Include figure files
\usepackage{dcolumn}% Align table columns on decimal point
\usepackage{bm}% bold math
\usepackage{color}

\usepackage[colorlinks,bookmarks=false,citecolor=blue,linkcolor=red,urlcolor=blue]{hyperref}
\def\be{\begin{equation}}       \def\ee{\end{equation}}
\def\bea{\begin{eqnarray}}      \def\eea{\end{eqnarray}}

\begin{document}
\title{Cooperation between electron-phonon coupling and electronic interaction in bilayer nickelates La$_3$Ni$_2$O$_7$}

\author{Jun Zhan}
\thanks{These authors equally contributed to the work.}
\affiliation{Beijing National Laboratory for Condensed Matter Physics and Institute of Physics, Chinese Academy of Sciences, Beijing 100190, China}
\affiliation{School of Physical Sciences, University of Chinese Academy of Sciences, Beijing 100190, China}

\author{Yuhao Gu}
\thanks{These authors equally contributed to the work.}
\affiliation{Beijing National Laboratory for Condensed Matter Physics and Institute of Physics, Chinese Academy of Sciences, Beijing 100190, China}
\affiliation{School of Mathematics and Physics, University of Science and Technology Beijing, Beijing 100083, China}

 \author{Xianxin Wu}\email{xxwu@itp.ac.cn}
 \affiliation{CAS Key Laboratory of Theoretical Physics, Institute of Theoretical Physics,
Chinese Academy of Sciences, Beijing 100190, China}

\author{Jiangping Hu}\email{jphu@iphy.ac.cn}
\affiliation{Beijing National Laboratory for Condensed Matter Physics and Institute of Physics, Chinese Academy of Sciences, Beijing 100190, China}
\affiliation{Kavli Institute for Theoretical Sciences, University of Chinese Academy of Sciences, Beijing 100190, China}
\affiliation{ New Cornerstone Science Laboratory, Beijing 100190, China}
\begin{abstract}
The recent observation of high-T$_c$ superconductivity in the bilayer nickelate La$_3$Ni$_2$O$_7$ under pressure has garnered significant interests. While researches have predominantly focused on the role of electron-electron interactions in the superconducting mechanism, the impact of electron-phonon coupling (EPC) has remained elusive. In this work, we perform first-principles calculations to study the phonon spectrum and electron-phonon coupling within La$_3$Ni$_2$O$_7$ under pressure and explore of the interplay between EPC and electronic interactions on the superconductivity by employing functional renormalization group approach. Our calculations reveal that EPC alone is insufficient to trigger superconductivity in La$_3$Ni$_2$O$_7$ under pressure. We identify unique out-of-plane and in-plane breathing phonon modes which selectively couple with the Ni $d_{z^2}$ and $d_{x^2-y^2}$ orbitals, showcasing an orbital-selective EPC. Within the bilayer two-orbital model, it is revealed that solely electronic interactions foster $s_{\pm}$-wave pairing characterized by notable frustration in the band space, leading to a low transition temperature. Remarkably, we find that this out-of-plane EPC can act in concert with electronic interactions to promote the onsite and interlayer pairing in the $d_{z^2}$ orbital, partially releasing the pairing frustration and thus elevating T$_c$. In contrast, the inclusion of in-plane EPC only marginally affects the superconductivity, distinct from the cuprates. Potential experimental implications in La$_3$Ni$_2$O$_7$ are also discussed.

\end{abstract}

\maketitle

The exploration of unconventional superconductivity in nickelates have been attracted considerable attentions, since the discovery of high-T$_c$ cuprates. In 2019, superconductivity with a transition tempeature (T$_c$) around 5-15 K was observed  in “infinite-layer" nickelates (Sr,Nd)NiO$_2$ thin films on a substrate ~\cite{Li2019,Osada2020,Pan2022,wang2022,ding_critical_2023}. In analogous to cuprates, the low-energy electronic structure of the 112 nickelate is dominated by the Ni $d_{x^2-y^2}$ orbital, where the additional Nd $d$ states around the Fermi level introduce a self-doping. The highest T$_c$ of 30 K can be achieved with external pressure~\cite{wang2022}. Last year, signatures of superconductivity with a T$_c\sim$ 80 K were reported in a new type of bilayer nickelates La$_3$Ni$_2$O$_7$ (LNO) within the Ruddlesden-Popper phase when the external pressure is greater than 14 GPa~\cite{sun2023}. The observation of zero resistance was independently confirmed~\cite{JGCheng2023,HQYuan2023,PhysRevX.14.011040-JGCheng}, with susceptibility measurements indicating filamentary superconductivity~\cite{LLSun2023}. With increasing pressure, the system undergoes a structural transition from the \textit{Amam} phase to \textit{Fmmm}/\textit{I4mmm} phase~\cite{sun2023,2023arXiv231109186W}. Distinct from cuprates, the formal valence is $d^{7.5}$ for Ni$^{2.5+}$ and the low-energy electronic structure is dominated by both $d_{x^2-y^2}$ and $d_{z^2}$ orbitals due to the presence of apical oxygens between layers~\cite{YaoDX,YZhang2023,Lechermann2023,Hirofumi2023possible,XWu,XJZhou2023,HHWen2023}. Superconductivity has also been reported in a trilayer nickelate La$_4$Ni$_3$O$_{10}$ under pressure~\cite{Kuroki2023T,HHWen2024T,JZhao2023T,YQi2023T,MWang2023T}, albeit with a lower T$_c$ of 20-30 K. Besides superconductivity, both bilayer and trilayer nickelates exhibit diverse correlated phenomena, including density wave transitions and non-Fermi liquid behaviors~\cite{sun2023,HQYuan2023,HHWen2023,chen2024electronic}. 

The mechanism of high T$_c$ superconductivity for the LNO nickelate have been intensively studied from the perspective of electron-electron interactions~\cite{Wang327prb, lu2024interlayer,HYZhangtype2,XWu,FangYang327prl,WeiLi327prl,Hirofumi2023possible,YifengYang327prb,YifengYang327prb2,YiZhuangYouSMG,tian2023correlation,Dagotto327prb,zhang2024structural,Jiang_2024,PhysRevB.108.L201121,liao2023electron,ryee2024quenched,luo2023hightc,fan2023superconductivity,KuWeiprl,KJiang:17402}, and a unified understanding remains elusive yet. 
Theoretical calculations reveal that the pressure-driven Lifshitz transition in the Fermi surfaces, characterized by the emergence of a hole pocket from the $d_{z^2}$ interlayer bonding state, is posited as a critical factor for superconductivity~\cite{sun2023}. Weak coupling analysis generates an $s_{\pm}$-wave pairing~\cite{Hirofumi2023possible,XWu,Wang327prb,FangYang327prl}. While, other studies highlight the critical role of the interlayer exchange coupling, Hund's rule coupling and orbital hybridization, advocating for the interlayer 
$s$-wave pairing ~\cite{lu2024interlayer,HYZhangtype2,WeiLi327prl}. Recently experimental measurements imply an indispensable role of electron-phonon coupling in determining electronic properties~\cite{talantsev2024debye,li2024ultrafast}. The crucial role of EPC in fostering superconductivity has already been demonstrated in both cuprates and iron based superconductors~\cite{Gunnarsson_2008}, especially FeSe/SrTiO$_3$ with an interface~\cite{LJJ2014,Wang_FeSESTO}. These, together with evidence of filamentary superconductivity, necessitate a thorough investigation into the role of EPC in LNO: questioning whether EPC alone can explain the observed high-T$_c$ superconductivity; and probing the specific characteristics of EPC within the bilayer structure, particularly its influence on correlated states in the context of its interplay with electronic interactions.

To elucidate the role of lattice degree of freedom in superconductivity of LNO,
our work delves into the phonon spectrum and electron-phonon coupling within the bilayer structure, with a particular focus on the intricate interplay between EPC and electronic interactions. Our calculations reveal that EPC in pressurized La$_3$Ni$_2$O$_7$ is insufficient to induce superconductivity. We identify out-of-plane and in-plane breathing phonon modes within the bilayer structure, which selectively couple with the Ni $d_{z^2}$ and $d_{x^2-y^2}$ orbitals, modeled through orbital-selective Su-Schrieffer-Heeger (SSH) type electron-phonon interactions. Within the bilayer two-orbital model, it is revealed that solely electronic interactions foster the $s_{\pm}$-wave pairing characterized by notable frustration in the band space, leading to a low transition temperature. Intriguingly, we find that the unique out-of-plane EPC can act in concert with electronic interactions to promote the onsite and interlayer pairing in the $d_{z^2}$ orbital. This partially releases the pairing frustration and thus elevates T$_c$. While, the inclusion in-plane EPC can hardly affect the superconductivity, distinct from the cuprates. Potential experimental implications in La$_3$Ni$_2$O$_7$ are also discussed.

\begin{figure}
\centerline{\includegraphics[width=0.5\textwidth]{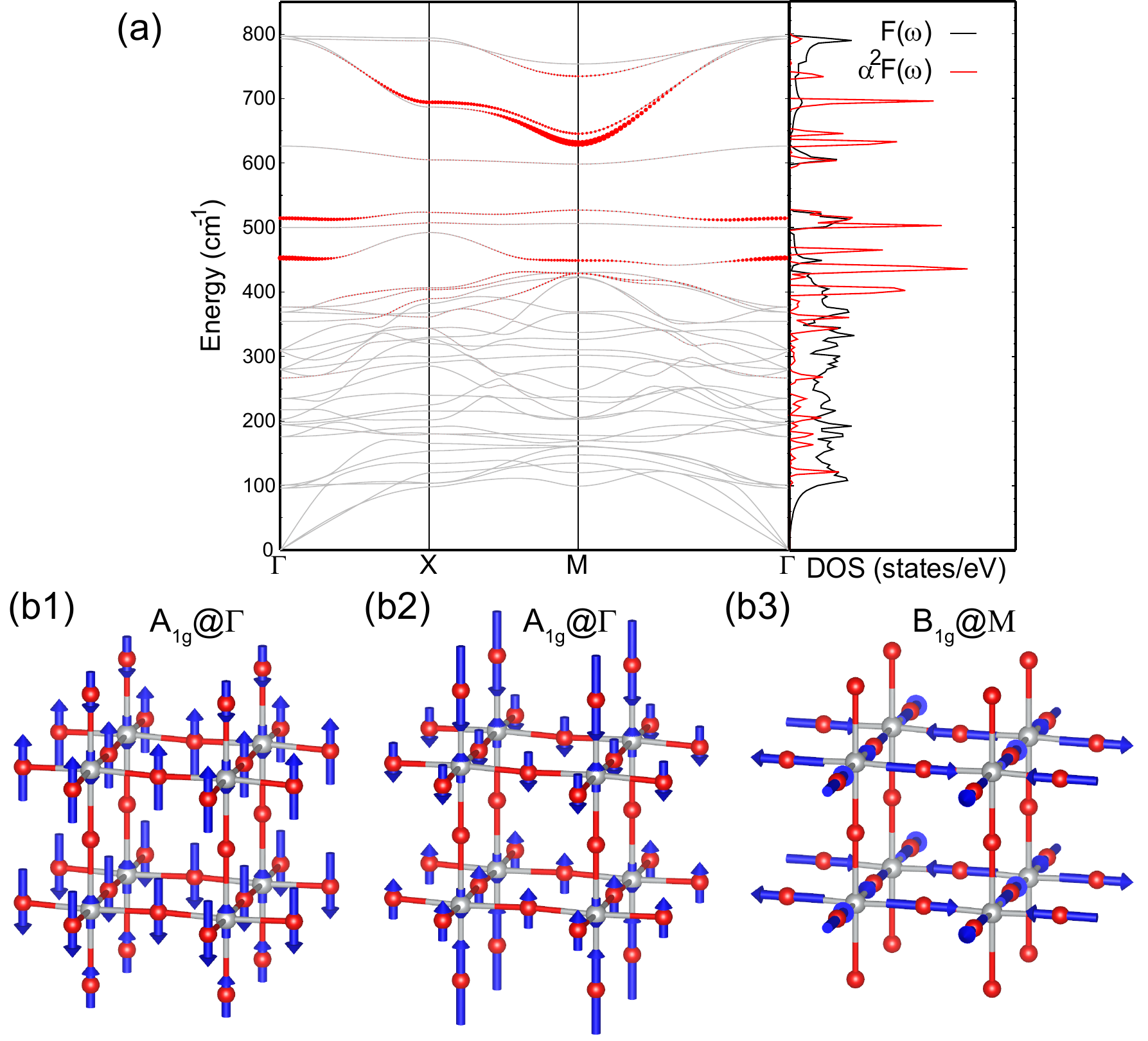}}
\caption{(color online) (a) Phonon spectrum, linewidth, density of states and Eliashberg spectral function of \ce{La3Ni2O7} with the space group {\it I4/mmm} at 30 Gpa. Out-of-plane phonon modes at the $\Gamma$ point with frequencies of 453 cm$^{-1}$ (b1) and 515 cm$^{-1}$ (b2) and in-plane phonon mode at the M point with a frequency of 630 cm$^{-1}$ (b3). \label{fig1}}
\end{figure}

{\it Electron-phonon coupling in DFT calculations}. Through density functional theory (DFT) calculations, we reveal the effect of electron phonon coupling on the superconductivity of La$_3$Ni$_2$O$_7$, with details of calculations provided in the supplementary materials (SM). This bilayer nickelate undergoes a structural transition with increasing pressure $P$ and exhibit superconductivity when $P$ is larger than 14 GPa~\cite{sun2023}. We choose a representative high-symmetry structure with \textit{I4/mmm} at 30 GPa~\cite{2023arXiv231109186W}. The phonon spectrum, linewidth, and density of states are illustrated in Fig.\ref{fig1}(a), where the circle sizes correspond to the phonon linewidth, highlighting regions of significant EPC. Notably, EPC predominantly affects one dispersive and two flat phonon bands. The dispersive band around 700 cm$^{-1}$ shows pronounced EPC near the M point, with the associated phonon mode depicted in Fig.\ref{fig1}(b3). This breathing mode, characterized by the $B_{1g}$ irreducible representation, involves in-plane oxygen atom motions within the Ni-O planes, in analogous to that in cuprates. The in-plane $\sigma$ bonding between Ni $d_{x^2-y^2}$ orbitals and O $p_{x,y}$ orbitals suggests its strong coupling with the Ni $d_{x^2-y^2}$ band, consistent with the prominent phonon linewidth in our calculations.
In contrast, two flat phonon bands around 500 cm$^{-1}$ demonstrate significant EPC near the $\Gamma$ point. These modes, associated with the $A_{1g}$ irreducible representation as shown in Fig.\ref{fig1}(b1) and (b2), involve predominantly out-of-plane motions of in-plane and apical oxygen atoms, and to a lesser extent, nickel atoms within the bilayer. The out-of-plane motion results in a weak in-plane dispersion. The pronounced interlayer bonding between Ni $d_{z^2}$ orbitals suggests that these out-of-plane breathing modes substantially influence Ni $d_{z^2}$ bands, especially when they cross the Fermi level at the high pressure. This mode is distinctive to the bilayer structure, lacking a counterpart in cuprates where the $d_{z^2}$ orbital remains inactive. 
Integrating the EPC spectral function $\alpha^2F(\omega)/\omega$, we obtain the total EPC constant $\lambda=0.11$ and the logarithmic averge frequency is $\omega_{ln}=495$ cm$^{-1}$. Employing a Coulomb pseudopotential $\mu^*=0.1$, the McMillan equation yields a superconducting transition temperature T$_c\approx 0$ solely from EPC. This clearly demonstrates that the main driving force of the observed superconductivity in La$_3$Ni$_2$O$_7$ is not EPC but rather electronic interactions. However, the role of EPC, especially when intertwined with electronic correlations, remains to be explored in the bilayer nickelates.

\begin{figure}
\centerline{\includegraphics[width=0.5\textwidth]{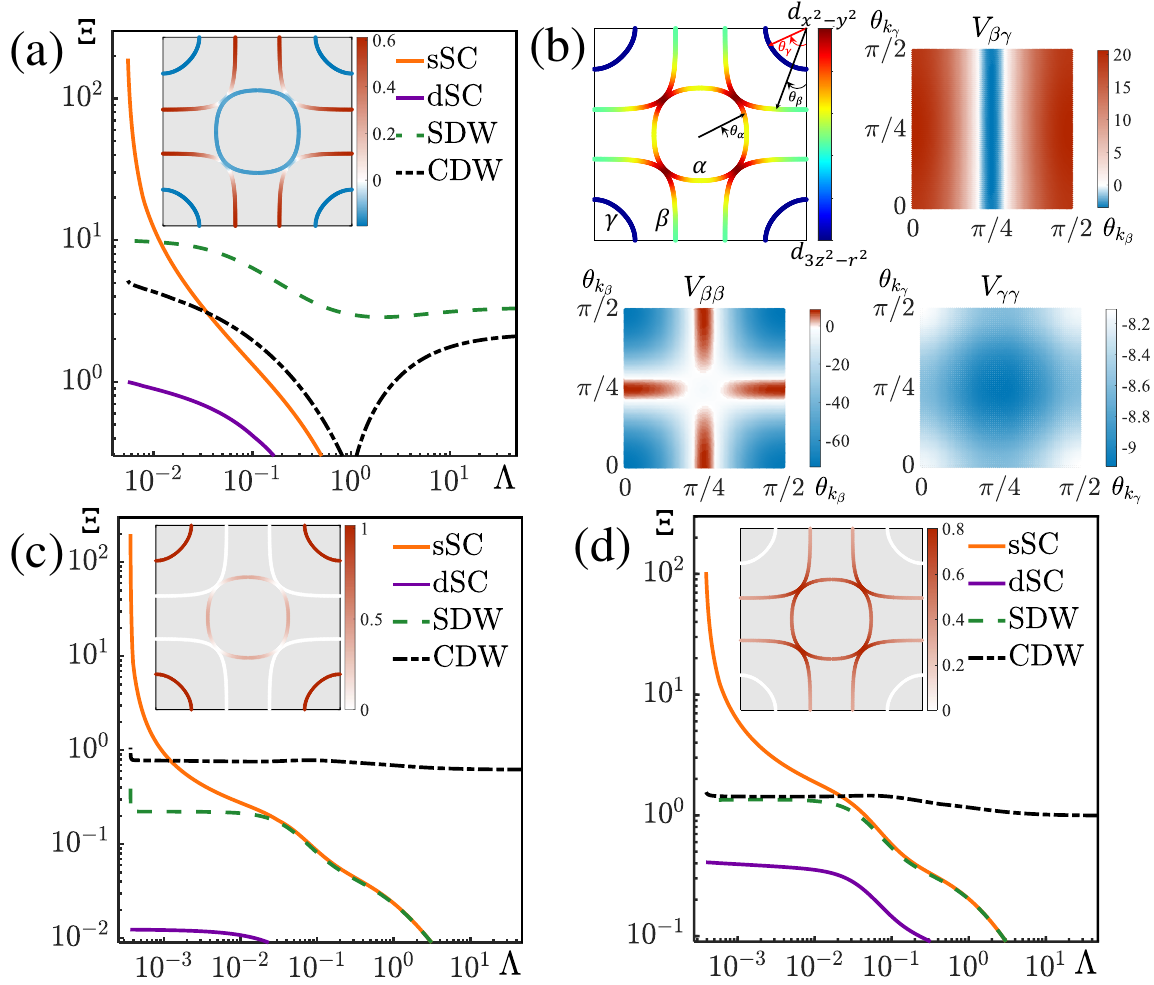}}
\caption{(color online). 
(a) RG flows of leading instabilities both particle-hole and particle-particle channels from sole electronic interactions with $U$=3 eV and $J/U=0.1$ and the inset shows the leading $s_{\pm}$-wave gap function. (b) Orbital characters on the Fermi surface (top left) and the effective pairing interaction in the band space. Here we use the angles $\theta_{\alpha,\beta,\gamma}$ to label momentum points on different Fermi surfaces. RG flows of leading instabilities from out-of-plane (c) and in-plane (d) EPC, where the EPC constant are $\lambda^z_\perp=0.5$ (c) and $\lambda^x_\parallel=0.8$ (d). Phonon frequency $\omega=0.05$ is used in
the calculations, unless explicitly specified. The insets display the gap function of leading pairings. \label{fig2}}
\end{figure}

{\it Pairing frustration from spin fluctuations and SSH electron phonon coupling}. 
The low-energy electronic structures of bilayer nickelates La$_3$NiO$_7$ are dominated by Ni $d_{x^2-y^2}$ and $d_{z^2}$ orbitals according to theoretical calculations and experimental measurements~\cite{YaoDX,YZhang2023,Lechermann2023,Hirofumi2023possible,XWu,XJZhou2023,HHWen2023}. At the high pressure, the low-energy physics can be described by a two-orbital tight-binding (TB) Hamiltonian~~\cite{YaoDX,YZhang2023,Lechermann2023,Hirofumi2023possible,XWu}, which reads,
\begin{equation}
    \mathcal{H}_0=\sum_{ij,\alpha\beta,\sigma}t_{\alpha\beta}^{ij}c_{i\alpha\sigma}^\dagger c_{j\beta\sigma}-\mu\sum_{i\alpha\sigma} c_{i\alpha\sigma}^\dagger c_{j\alpha\sigma}.
\end{equation}
Here $i,j=(m,l)$ labels the in-plane lattice site ($m$) and layer index ($l=t,b$), $\sigma$ labels spin, and $\alpha,\beta=x,z$ labels the Ni orbitals with $x$ denoting the $d_{x^2-y^2}$ and $z$ the $d_{z^2}$ orbital. $\mu$ is the chemical potential and the adopted parameters and resulted band structures are provided in SM. There are three Fermi surfaces as illustrated in Fig.\ref{fig2} (b): the $\alpha$ electron pocket arises from the interlayer bonding state of $d_{x^2-y^2}$ and  $d_{z^2}$ orbital and the hole-like $\beta$ and $\gamma$ pockets originate from the interlayer anti-bonding state of $d_{x^2-y^2}$ and bonding state of $d_{z^2}$ orbital, respectively. For the electronic interactions (EI), we adopt the general multi-orbital Hubbard interactions,
\begin{equation}
    \begin{aligned}
    \mathcal{H}_{\mathrm{EI}}&= \sum_{i\alpha}Un_{i\alpha\uparrow}n_{i\alpha\downarrow}+\sum_{i,\alpha\neq \beta}J_Pc_{i\alpha\uparrow}^{\dagger}c_{i\alpha\downarrow}^{\dagger}c_{i\beta\downarrow}c_{i\beta\uparrow}\\
    &+\sum_{i,\alpha< \beta,\sigma\sigma^{\prime}}(U^{\prime}n_{i\alpha\sigma}n_{i\beta\sigma^{\prime}}+J_Hc_{i\alpha\sigma}^{\dagger}c_{i\beta\sigma}c_{i\beta\sigma^{\prime}}^{\dagger}c_{i\alpha\sigma^{\prime}}),
    \end{aligned}
\end{equation}
where $U/U^{\prime}$ is the intra-orbital/inter-orbital Hubbard repulsion, $J_{H}$ is the Hund’s coupling, and
$J_P$ is the pair-hopping interaction. The standard Kanamori relations $U = U^{\prime} + 2J_{H}$ and $J_H=J_P$ are adopted in this work.
The nature of superconductivity has been studied from both weak-coupling and strong-coupling perspective, yet a consensus remains elusive. Here, we adopted the functional renormalization group (FRG) approach, treating all particle-hole and
particle-particle channels on equal footing, thereby providing a nuanced depiction of correlated states spanning from weak to intermediate coupling regimes~\cite{FWang2010,Metzner2012,Platt2013}. The details are provided in SM. Utilizing the representative parameters $U=3$ eV and $J/U=0.1$, the RG flows of leading eigenvalues $\Xi$ in different channels with decreasing cutoff $\Lambda$ are illustrated in Fig.\ref{fig2}(a). A dominant $s_{\pm}$-wave superconductivity, featuring the sign-reversed gaps on the bonding and anti-bonding states, emerges at a low scale $\Lambda_c =5.5\times 10^{-3}$. The spin and charge density wave (S/CDW) and $d$-wave pairing are subdominant. The sign change of the $s_{\pm}$ gap ( inset of Fig.\ref{fig2}(a) ) is derived from the effective repulsive inter-pocket interaction $V_{\beta\alpha/\beta\gamma}$ from the spin fluctuations~\cite{Wang327prb,XWu}, as shown in Fig.\ref{fig2}(b). Moreover, the strong intra-pocket Cooper pairing scattering on the $\beta$ pocket leads to a large superconducting gap on this pocekt. In the real space, the pairing is dominated in the intra-orbital onsite ($\Delta_{ll,0}^{\alpha\alpha}$) and inter-layer ($\Delta_{l\bar{l},0}^{\alpha\alpha}$) channels of the two orbitals. For the $d_{x^2-y^2}$ orbital, the onsite and interlayer pairings have the same sign $\Delta^{xx}_{ll,0}=-0.14$ and $\Delta^{xx}_{l\bar{l},0}=-0.04$. In contrast, the $d_{z^2}$ orbital exhibits a significantly larger pairing gap, with a sign reversal between onsite and interlayer pairings, $\Delta^{zz}_{ll,0}=0.42$ and $\Delta^{zz}_{l\bar{l},0}=-0.51$. These features result in an enhanced gap on the anti-bonding $\beta$ bands but a diminished gap on the bonding $\alpha,\gamma$ bands, aligning with the analysis of effective interactions. Given the substantial density of states carried by the $\gamma$ pocket, which nonetheless acquires a small superconducting (SC) gap, this mismatch between SC gaps and DOS distribution induces pairing frustration, leading to a relatively low transition temperature of $T_c \sim 10^{-3}$.

We further study the effect of sole EPC on superconductivity based on FRG calculations~\cite{Wang_FeSESTO,Wang_Hostein,Wang_SSH}. To model the two prominent phonon modes with strong EPC in our DFT calculations, we consider the SSH type electron phonon interaction in the two-orbital model, which describes phonons selectively coupled to electron hopping in different orbitals. The Hamiltonian of phonon and EPC interaction reads
\begin{equation}
		\begin{aligned}
	\mathcal{H}_{\mathrm{EPC}}=\omega\sum_{\langle ij \rangle}b_{ij}^\dagger b_{ij}
        +\sum_{\langle ij \rangle\sigma} g_{ij}^{\alpha} (b_{ij}^\dagger+b_{ij})(c_{i\alpha\sigma}^\dagger c_{j\alpha\sigma}+\mathrm{H.c.}),
		\end{aligned}
\end{equation}
with $\langle ij \rangle$ denoting the nearest-neighbor (NN) in-plane and out-of-plane sites. Here Einstein phonons with a frequency of $\omega$ are adopted and the $b^{\dagger}_{ij=\parallel,\perp}$ creates an optical phonon on in-plane or vertical NN bond $ij$, which couples to the electrons in the $d_{x^2-y^2,z^2}$ orbital with the coupling strength $g_{ij}^{\alpha}$. To study the effect of EPC on the correlated states, we integrate out the phonon degrees of freedom and obtain a retarded electron-electron attraction,
	\begin{eqnarray}
		  &V_{\mathrm{eff}}=\frac12 \sum_{\langle ij \rangle \nu \alpha }\Pi_{ij}^{\alpha}(\nu)B_{ij}^{\alpha}(\nu)B_{ij}^{\alpha}(-\nu), 
%            &\quad B_{ ij}(\nu)=\sum_{\sigma}\int_{0}^{\beta} e^{i\nu\tau}(c^{\dagger}_{ia\sigma}(\tau)c_{ja\sigma}(\tau) + \mathrm{H.c.}),
	\end{eqnarray}
with $\Pi_{ij}^{\alpha}(\nu)=-\frac{\lambda_{ij}^{\alpha}}{N_F}\frac{\omega^2}{\nu^2+\omega^2}$ and $B_{ ij}(\nu)=\sum_{\sigma}\int_{0}^{\beta} e^{i\nu\tau}(c^{\dagger}_{i\alpha\sigma}(\tau)c_{j\alpha\sigma}(\tau) + \mathrm{H.c.})$. Here $\nu$ is bosonic Mutsubra frequency, 
 $N_F$ is the DOS at the Fermi level and $\lambda_{ij}^{\alpha}=2N_F(g_{ij}^\alpha)^2/(\hbar \omega)$ is the dimensionless EPC constant. With this effective electron interaction, we perform FRG calculations to explore the instabilities from two kinds of EPC. The typical RG flows for $\lambda^{z}_{\perp}=0.5$ and $\lambda^{x}_{\parallel}=0.8$ are displayed in Fig. \ref{fig2} (c) and Fig. \ref{fig2} (d), respectively. Apparently, in both cases, charge fluctuations are promoted and EPC generate an $s$-wave pairing without any sign change, in contrast to electronic interactions. The gap functions in two cases, however, are distinct: the vertical EPC favors a large isotropic SC gap on the $\gamma$ pocket but the in-plane EPC favors large anisotropic SC gaps on $\alpha,\beta$ pockets. These features are intimately related to the orbital-selective EPC and orbital characters on the Fermi surfaces. Strong attractive interactions are generated in the $d_{z^2}$ and $d_{x^2-y^2}$ orbitals in the former and latter cases and this results a large SC gap on the $d_{z^2}$-dominant $\gamma$ pocket and on the $\alpha,\beta$ pockets dominated by the $d_{x^2-y^2}$ orbital along the diagonal direction. The effective pairing interactions are shown in SM, consistent with our analysis here. It is noteworthy that the transition temperatures remain relatively low despite employing sizable EPC constants. In addition, superconductivity vanishes when the EPC constants are set to values derived from DFT calculations.

\begin{figure}[t]
\centerline{\includegraphics[width=0.5\textwidth]{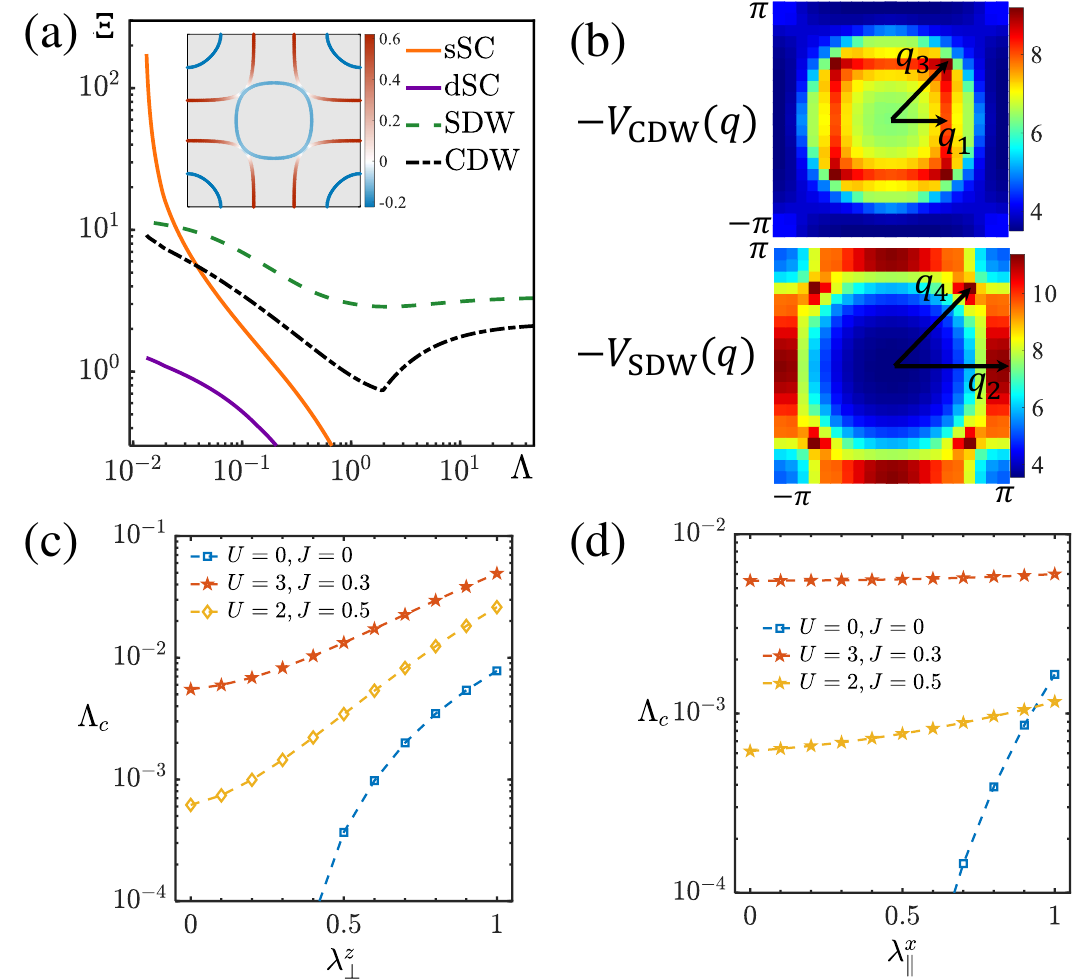}}
\caption{(color online). (a) Typical RG flows of leading instabilities with electronic interactions and out-of-plane EPC ($U$=3 eV, $J/U=0.1$ and $\lambda^z_{\perp}=0.5$) and the inset shows the leading gap function.  (b) Leading momentum-dependent eigenvalues of the renormalized interactions in the CDW and SDW channels. Critical cutoffs of the leading $s_{\pm}$-wave pairing as a function of out-of-plane (c) and in-plane (d) EPC constants for various interaction parameters. The transition temperatures increase almost exponentially with the out-of-plane EPC constant.    \label{fig3}}
\end{figure}

{\it Cooperation between out-of-plane EPC and electronic correlation on superconductivity}. Given that either electronic interactions or EPC results in low SC transition temperatures within the bilayer two-orbital model, we explore the combined effect of these two types of interactions on superconductivity. Fig.\ref{fig3} (a) illustrates the representative RG flow with electronic interactions and out-of-plane EPC. We observe that both spin and charge fluctuations increase rapidly with the decreasing cutoff for  $\Lambda<1$. Fig.\ref{fig3}(b) displays the leading negative eigenvalues of the renormalized interactions $V(\bm{q})$ in the CDW and SDW channels at the critical cutoff. By introducing the out-of-plane EPC, the peaks at the nesting vectors $\bm{q}_2$ and $\bm{q}_4$ in the SDW channel gets slightly enhanced, and the peak at $\bm{q}_3=(\pi/2,\pi/2)$ attributed to the $\gamma$ pocket in the CDW channel become prominent. These fluctuations significantly promote the $s_{\pm}$-wave pairing and thus it diverges at $\Lambda_c=2.2\times10^{-2}$, almost four times the value from sole electronic interactions.  As shown in the inset of Fig.\ref{fig3} (a),  the gap amplitudes across the entire $\gamma$ pocket and on the $\beta$ pocket around the X/Y point get enhanced,
which releases the pairing frustration from EI and thus leads to an enhanced transition temperature. We further perform calculations for various interaction parameters as a function of EPC constant $\lambda^z_\perp$, as shown in Fig.\ref{fig3}(c). The leading pairing state is consistently $s_{\pm}$-wave and the transition temperature exhibits an almost exponential increase with the EPC constant $\lambda^z_\perp$. This suggest a synergistic effect between out-of-plane EPC and electronic interactions in promoting superconductivity. However, the in-plane EPC coupling on the $d_{x^2-y^2}$ orbital will slightly modify the SDW fluctuations at $\bm{q}_4 \approx (3\pi/4,3\pi/4)$ and hardly change the SC transition temperature of $s_{\pm}$-wave pairing, as demonstrated in Fig.\ref{fig3} (d). These results reveal that in the bilayer model, different phonon modes coupled to electrons exert varying impacts on superconductivity, reminiscent of the behavior observed in cuprates~\cite{Gunnarsson_2008}.

\begin{figure}[t]
\centerline{\includegraphics[width=0.5\textwidth]{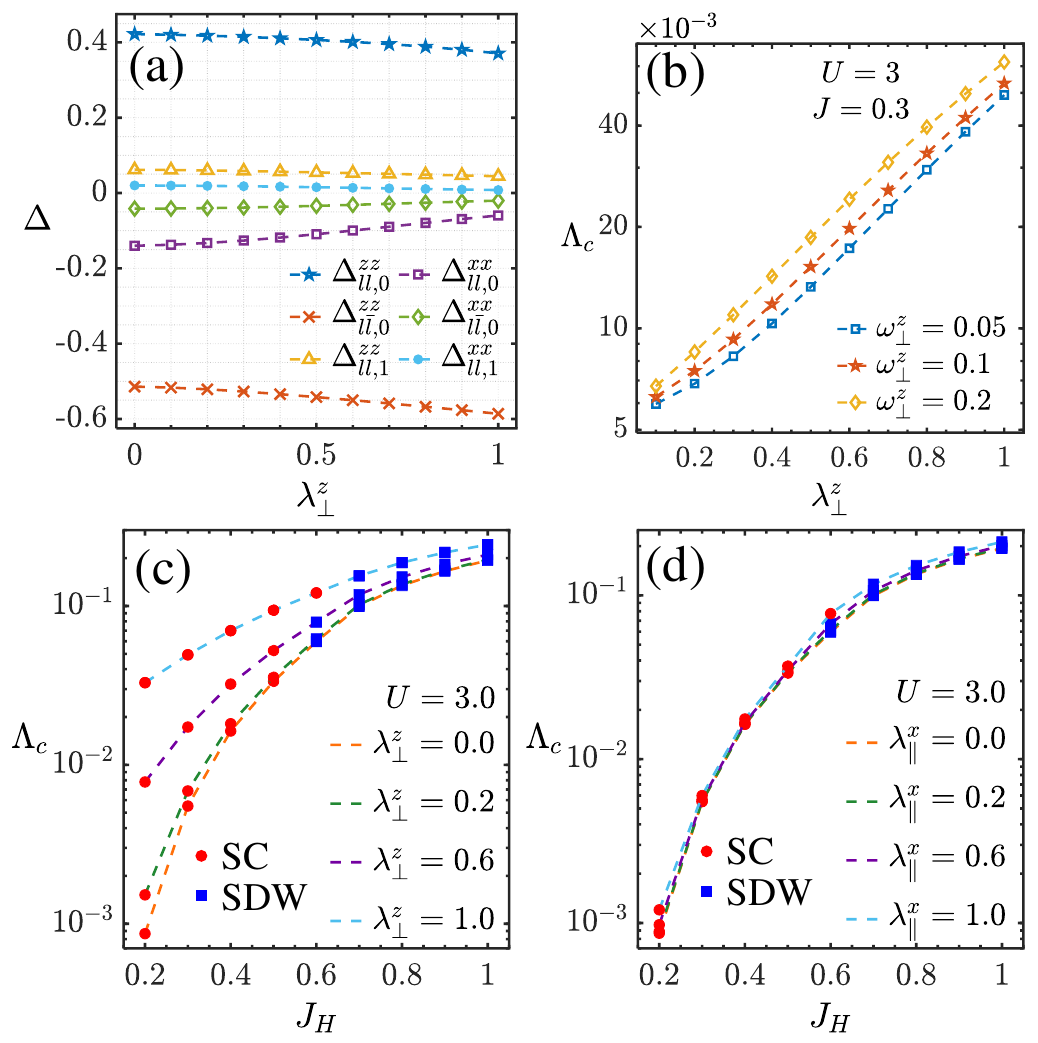}}
\caption{(color online). (a) Real space pairing components change as a function of out-of-plane EPC constant $\lambda_{\perp}^{z}$. (b)  Critical cutoffs
of the SC as a function of out-of-plane EPC constant $\lambda_{\perp}^{z}$ with different phonon frequencies. Leading electronic instabilities and critical cutoff a function of $J_H$ with different out-of-plane (c) and in-plane (d) EPC constants. \label{fig4}}
\end{figure}

{\it Impact of EPC on real-space pairing and SDW states}.  We further analyze the effect of EPC on SC pairing in the real space and SDW states. Through the analysis of pairing vertex in the real space, we can obtain the dominant pairing component up to the NN sites in two orbitals and the real-space SC parameter $\Delta^{\alpha\beta}_{ll',\delta}$ denotes the spin-singlet paring between the $\alpha$ orbital in the layer $l$ and $\beta$ orbital in layer $l'$ with an in-plane distance $\delta$=0 (onsite), 1 (NN). In the case with both EI and out-of-plane EPC, the SC components as a function of $\lambda^z_\perp$ are illustrated in Fig.\ref{fig4}(a). With increasing EPC constant, the dominant gap amplitudes on the $d_{z^2}$ orbitals decreases in the onsite channel but increase in the interlayer channel. While, both onsite and interlayer SC gaps on the $d_{x^2-y^2}$ orbital get reduced and the NN pairings in two orbitals also decrease. After integrating the out-of-plane SSH phonon, onsite and interlayer attractive interactions are generated in the $d_{z^2}$ orbital, leading to an onsite and NN pairing with the same sign in the mean-field level. When these pairings are added to the interaction-driven sign-reversed pairing in the $d_{z^2}$ orbital, the interlayer pairing increases but the onsite pairing decreases. This explains the observed variation of pairing in the $d_{z^2}$ orbital with increase EPC constant. The interference between EI and out-of-plane EPC will also suppress the pairing in the $d_{x^2-y^2}$ orbital. As the pairing of the (anti-)bonding band is the summation (difference) between the onsite and interlayer pairing, the variation of real-space component leads to enhanced SC gaps on the bonding $\alpha,\gamma$ bands and 
 suppressed the SC gap on the anti-bonding $\beta$ band
. We also study the variation of transition temperature for various phonon frequencies, as shown in Fig.\ref{fig4}(b). With the increase in frequency, the effective attraction from phonons also grows, leading to a rise in T$_c$ enhancement. This observation is in line with the phonon-induced isotope effect. In addition, we explore the impact of EPC on the SDW state as a function of Hund's rule coupling $J_H$. The obtained critical cutoffs for different out-of-plane and in-plane EPC constants are displayed in Fig.\ref{fig4} (c) and (d), respectively. For a fixed onsite repulsion, increasing $J_H$ drives a transition from superconductivity to SDW with a wave vector of $\bm{q}_4$. It is evident that the out-of-plane EPC more significantly promotes both superconductivity and SDW and the T$_c$ enhancement  decreases with increasing $J_H$. While, even a very strong in-plane EPC results in only a marginal increase in the T$_c$ of superconductivity.

{\it Discussion and conclusion}. 
The out-of-plane breathing phonon within the bilayer structure and the associated EPC have dramatic effects on superconductivity and other correlated states. So far, the mechanism of superconductivity in La$_3$Ni$_2$O$_7$ under pressure is still controversial. EPC alone cannot induce superconductivity in La$_3$Ni$_2$O$_7$ according to our DFT calculations. In the weak to intermediate coupling regime, the interaction-driven SC transition temperature is low in the bilayer two-orbital model due to the induced pairing frustration. The further inclusion of out-of-plane EPC can release this frustration and significantly enhance T$_c$. By adopting the out-of-plane EPC constant $\lambda=0.1\sim0.2$ , the T$_c$ enhancement relative to the EI case is about 10\% $\sim$ 40\%. Under pressure, the separation between two Ni-O layer decreases and this can induce phonon hardening and significantly enhance the out-of-plane EPC, thereby elevating T$_c$. These may provide an explanation for the observed superconductivity in La$_3$Ni$_2$O$_7$ with an extraordinary high T$_c$ of 80 K under high pressure. The proposed scenario can be examined in the experiments. The identified out-of-plane breathing phonon can be detected in Ramman and neutron scattering measurements. The EPC will also affect the electron spectrum and induce kink features in the dispersion, which can be detected in ARPES measurement. If the EPC is partially responsible for the superconductivity, we expect oxygen-isotope effect in La$_3$Ni$_2$O$_7$. The EPC also promotes the CDW and SDW fluctuations and especially the enhanced CDW at $\bm{q}_3=(\pi/2,\pi/2)$ is consistent with recent RIXS measurements~\cite{chen2024electronic}. In both EI and EPC, the $d_{z^2}$ orbital plays a pivotal role in the correlated states according to our analysis, marking a distinction from cuprates.

In conclusion, our calculations reveal that EPC alone is insufficient to induce superconductivity in pressurized La$_3$Ni$_2$O$_7$. However, when EPC is combined with electronic interactions, it significantly enhances superconductivity, leading to a relatively high transition temperature. Notably, we identify critical out-of-plane and in-plane breathing phonon modes within the bilayer structure, which selectively couple with the Ni $d_{z^2}$ and $d_{x^2-y^2}$ orbitals, demonstrating an orbital-selective EPC. Within the bilayer two-orbital model, it is revealed that solely electronic interactions foster $s_{\pm}$-wave pairing with notable frustration, leading to a relative low T$_c$. Remarkably, our study reveals that the unique out-of-plane EPC can act in concert with electronic interactions to significantly boost $s_{\pm}$-wave superconductivity, thereby elevating T$_c$. Our study highlights the indispensable role of electron-phonon coupling in conjunction with electronic correlations in fostering superconductivity, offering fresh and valuable insights into the mechanisms underpinning high T$_c$ superconductivity in bilayer nickelates under pressure.

{\it Acknowledgments}. We acknowledge the supports by the Ministry of Science and Technology (Grant No. 2022YFA1403901), National Natural Science Foundation of China (No. 11920101005, No. 11888101, No. 12047503, No. 12322405, No. 12104450) and the New Cornerstone Investigator Program. X.W. is supported by the National Key R\&D Program of China (Grant No. 2023YFA1407300) and the National Natural Science Foundation of China (Grant No. 12047503). Yuhao Gu also acknowledges the supports from China Postdoctoral Science Foundation Fellowship (No.2023T160675).

{\it Note added}. During the preparation of this work, we became aware of two independent
studies of phonon properties and electron-phonon coupling in  La$_3$Ni$_2$O$_7$ under pressure~\cite{yi2024antiferromagnetic,ouyang2024absence}. Their conclusion that the sole EPC is insufficient to drive superconductivity in La$_3$Ni$_2$O$_7$ is consistent with ours. Additionally, we extends the analysis to the combined effect of EPC and EI on superconductivity.

\bibliography{references_new0831}% Produces the bibliography via BibTeX.
 \bibliographystyle{apsrev4-1}

\end{document}